\documentclass[preprint]{elsarticle}
\usepackage{amssymb}
\usepackage{natbib}

\def\eslt{\not\!\!{E_T}}

\begin{document}

\begin{frontmatter}

\title{Measuring the Higgsplosion Yield: Counting Large Higgs Multiplicities
at Colliders\tnoteref{t1}}
\tnotetext[t1]{UH-511-1277-2017}

\author[jsg]{James~S.~Gainer}
\ead{jgainer@hawaii.edu}
\address[jsg]{Dept. of Physics and Astronomy, University of Hawaii, Honolulu, HI 96822, USA}

\begin{abstract}
Recent work has brought renewed attention to the possibility that
the cross section for producing $n$ Higgs bosons grows large with $n$
at a sufficiently energetic hadron collider.  In particular, this
``Higgsplosion'' mechanism has been suggested as a solution to the hierarchy
problem.
We investigate the phenomenology of this scenario.  Discovery is
trivial, so we consider several variables for use in measuring
large Higgs multiplicities and evaluate their effectiveness.
We find that a $10\%$ level measurement of the number of Higgs bosons
is possible with a handful of events, using the scalar sum of jet
transverse momenta, but determining the exact multiplicity
may take a few thousand events, depending on the degree of statistical
significance desired for the measurement.
While this situation may be acceptable given the potentially large
cross sections in this scenario, future research
to improve the measurement is warranted.
\end{abstract}

%

\end{frontmatter}

\section{Introduction}

The cross section for producing $n$ Higgs bosons
appears to grow exponentially large in perturbative calculations due to the
scaling of the number of diagrams (which goes as $n!$).
This fact has long been noted and has been much discussed~\cite{Cornwall:1990hh,
Goldberg:1990qk, Voloshin:1992mz, Brown:1992ay, Voloshin:1992rr, Voloshin:1992nu,
Smith:1992rq, Voloshin:1992xb, Argyres:1992np, Gorsky:1993ix, Libanov:1994ug,
Voloshin:1994yp, Libanov:1995gh, Son:1995wz, Libanov:1997nt, Khoze:2014kka,
Jaeckel:2014lya, Khoze:2015yba, Degrande:2016oan, Khoze:2017tjt, Voloshin:2017flq}.
It is at present unclear whether this apparent scaling reflects reality or
an incomplete calculation; it could also be that this scaling of the cross section
with $n$, which is calculated from only the interactions of the scalar particles,
is violated by loops involving heavy fermions~\cite{Voloshin:2017flq}.

Recently the asymptotic behavior of the multiple Higgs production cross section
with respect to $n$ has been invoked as a solution to the hierarchy
problem~\cite{Khoze:2017tjt}.
In this work, the production of a large number of Higgs bosons is referred to as
``Higgsplosion''.
This work also, importantly, pointed out a ``Higgspersion'' mechanism, via which
the exponential growth in Higgs cross sections is tamed via the momentum-dependent
width for a off-shell Higgs decaying to $n$ Higgs bosons becoming large.
This prevents the amplitude from violating unitarity at some sufficiently large value
of $n$.
Thus Higgsplosion is a viable possibility, though, as noted above,
by no means a definite prediction.

Ref.~\cite{Khoze:2017tjt} suggests that Higgsplosion may also solve the hierarchy problem.
The argument goes roughly as follows.  Any new heavy particle which couples to the
Higgs has a large partial width for decay to $n$ Higgses.  This large width cuts off
integrals involving the heavy particle; therefore quadratic corrections to the Higgs
mass from new heavy particles are cut off at a scale far below the Planck scale.
Ref.~\cite{Khoze:2017tjt} suggests this scale may be as low as $\sim 25$ TeV.
This interesting behavior is reminiscent of
classicalization~\cite{Dvali:2010jz, Dvali:2010ns, Dvali:2016ovn}
with the $n$-Higgs boson state playing the role of the classicalon.

As far as we are aware, the phenomenological consequences of the Higgsplosion
scenario have not been considered in detail.
Yet the exoticness of the scenario raises interesting questions.
Events with $\mathcal{O}(100-200)$ Higgs
bosons, which we will term ``Higgsplosion events'' in the remainder of this letter,
are easy to detect.  For example, none of our Monte Carlo simulated Higgsplosion events
had fewer than $20$ jets with $p_T > 50$ GeV; in general the jet multiplicities
are even larger and many of these jets have much higher $p_T$.  We therefore do not
need to consider standard model (SM) backgrounds.  In the event that Higgsplosion
is realized in nature, discovery would only require enough events for us to
convince ourselves that some strange detector malfunction were not to blame.

We therefore turn
our attention to what we can measure about Higgsplosion events.  The most
obvious property is the number, $n$, of Higgs bosons produced in the event.
This measurement proves to be not entirely straightforward, as
the large number of Higgs bosons, and consequently of decay products,
provide challenges that would be absent for a small or intermediate
Higgs multiplicity.
In the remainder of this
letter we will examine variables that could be used to measure
the Higgs multiplicity in Higgsplosion events.
Future research on this issue should consider potential improvements from
(a) multivariate analyses and (b) the use of variables related to jet substructure
to reduce the number of events needed for this measurement.

\section{Procedure}

We will consider three benchmark scenarios, namely
$150$ or $200$ Higgs bosons at a $100$ TeV collider and
$100$ Higgs bosons at the $14$ TeV CERN Large Hadron Collider (LHC).
While Ref.~\cite{Khoze:2017tjt} does not suggest that the $100$-Higgs boson events will
be visible at the LHC, we consider this scenario to have something relevant
to the current experimental situation.  (Also the ``theory error bars'' are large
in these suggestive arguments about the large $n$ Higgs cross sections.)
Ref.~\cite{Khoze:2017tjt} suggests that cross sections reach observable levels for
$n \gtrsim$ 130 at future colliders; so $n = 150$ and $n = 200$
should be seen as the more realistic scenarios.

In each of these scenarios we will follow the approach of Higgsplosion calculations
and consider Higgs bosons as being produced at threshhold.
This is sensible also because the exponential scaling of the $n$-Higgs production
cross section with $n$ means that producing an extra Higgs is generally favored
over having additional phase space available, up to the point where
suppressions from parton distribution functions (pdfs) or Higgspersion
cuts off the cross section.

We construct the ``parton level'' events explicitly by
writing Les Houches Accord Event(LHE) files~\cite{Alwall:2006yp}
with $n$ Higgs bosons at rest with respect to
each other.  Each event has an overall value of longitudinal momentum found from
the distribution we calculate using the \verb^NNPDF23_lo_as_0130_qed^ pdf~\cite{Ball:2014uwa}
as implemented in LHAPDF~\cite{Buckley:2014ana}.
We note, however that even at $100$ TeV, since $p_Z \sim 20 - 50$ GeV for
individual Higgs bosons in the event,
the Higgses are non-relativistic in the lab frame and the effect of this
longitudinal boost (or likewise a transverse boost from ISR) will be minimal.
For this reason the choice of collider in the event simulation is essentially
irrelevant.

We pass the LHE file to Pythia8~\cite{Sjostrand:2014zea} and the resulting HepMC~\cite{Dobbs:2001ck} file to DELPHES 3,
which contains FastJet~\cite{Cacciari:2011ma} for jet algorithms,
where we use the DELPHES 3 card described in Ref.~\cite{Baer:2016wkz},
but with leptons considered isolated if the activity in the detector
within a $\Delta R < 0.2$ cone is less than the minimum of $5$ GeV and $0.15$ times
the lepton $p_T$.
(While we provide the details of our simulation in the
interest of transparency,
we do not expect the details of our detector simulation
to have a significant qualitative effect on our results.)
We then use the LHCO files produced by
DELPHES to calculate various observables.

\section{Results}

\begin{figure}[ht]
\begin{center}
\includegraphics[width = \textwidth]{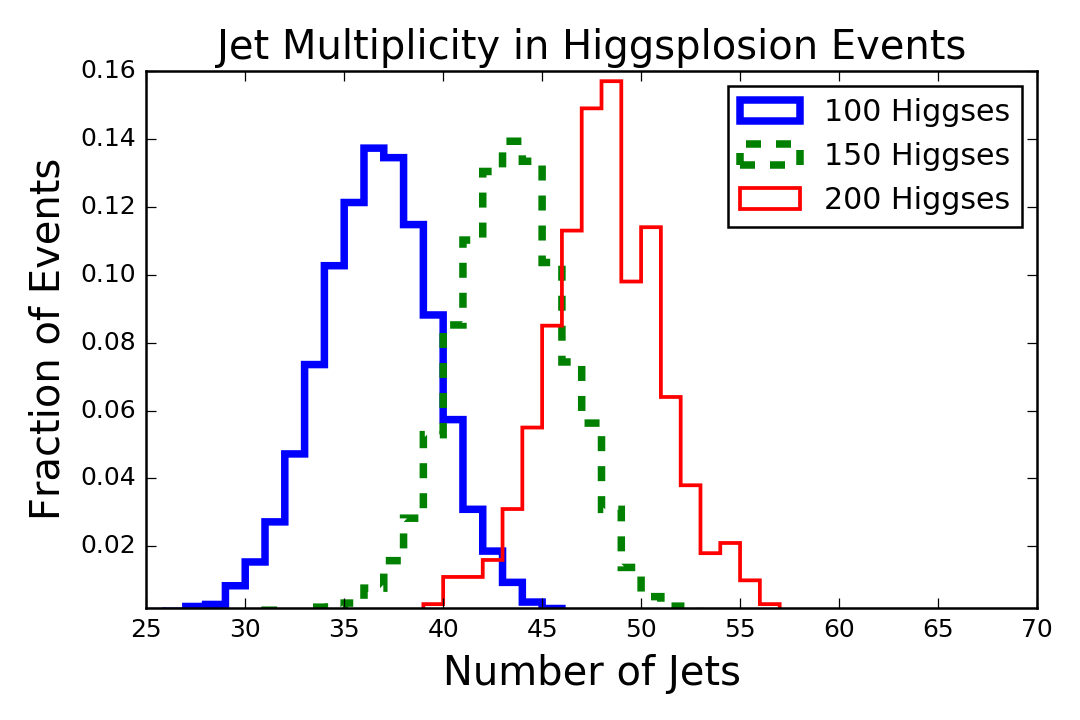}
\caption{Multiplicity of $p_T > 50$ GeV anti-$k_T$ jets
with $\Delta R < 0.4$ for $100$, $150$, and $200$
Higgs events in the final state.}
\label{ref:jet_multiplicity}
\end{center}
\end{figure}

\begin{figure}[ht]
\begin{center}
\includegraphics[width = \textwidth]{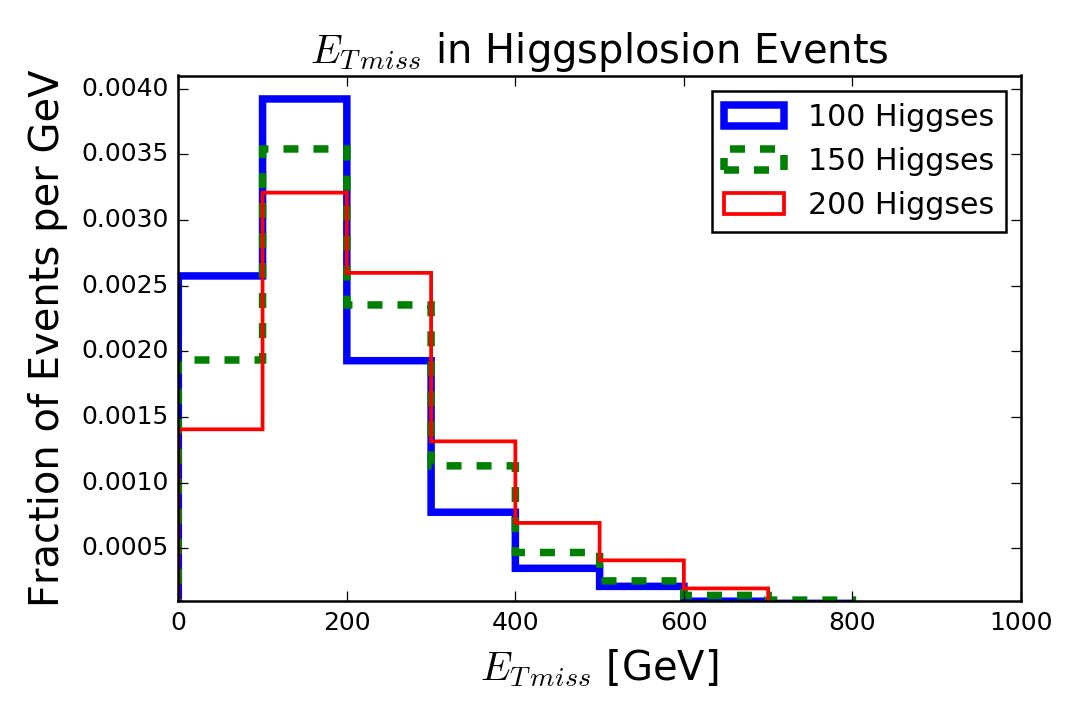}
\caption{Missing transverse energy distribution
for $100$, $150$, and $200$
Higgs events in the final state.}
\label{fig:MET}
\end{center}
\end{figure}

We use our simulated data to determine which variables will be most useful
for measuring the multiplicity of Higgs bosons in Higgsplosion events.
We first consider the number of jets in the final state,
which is shown in Figure~\ref{ref:jet_multiplicity}.
Naturally, the jet multiplicity is proportional to the Higgs multiplicity,
though the coefficient of proportionality is much less than one.
For example, we find that the mean number of jets for a Higgs multiplicity
of $100$ is $36.2$, while the mean jet multiplicity for $150$ Higgs events
is $42.8$.  Clearly decay products from many different Higgs bosons
may contribute to the same jet.
We do not consider $b$-tagging here because many jets
in a given event include decay products from multiple high $p_T$ $b$ quarks.  We feel a detailed analysis of
$b$-tagging efficiencies in this situation, which is beyond the scope of this letter,
would be required to draw even preliminary conclusions about the multiplicity of
$b$-tagged jets.
We do consider the number of isolated leptons, but find that it is a poor variable.
Due to the isolation requirements the number of leptons observed actually decreases
slightly with the increase in jet (and Higgs) multiplicity.
Another variable of limited efficacy is the missing transverse momentum ($\eslt$),
which we plot in Figure~\ref{fig:MET}.
We see that this variable increases weakly with the increase in Higgs multiplicity,
and, in general, takes on relatively low values.

\begin{figure}
\begin{center}
\includegraphics[width = \textwidth]{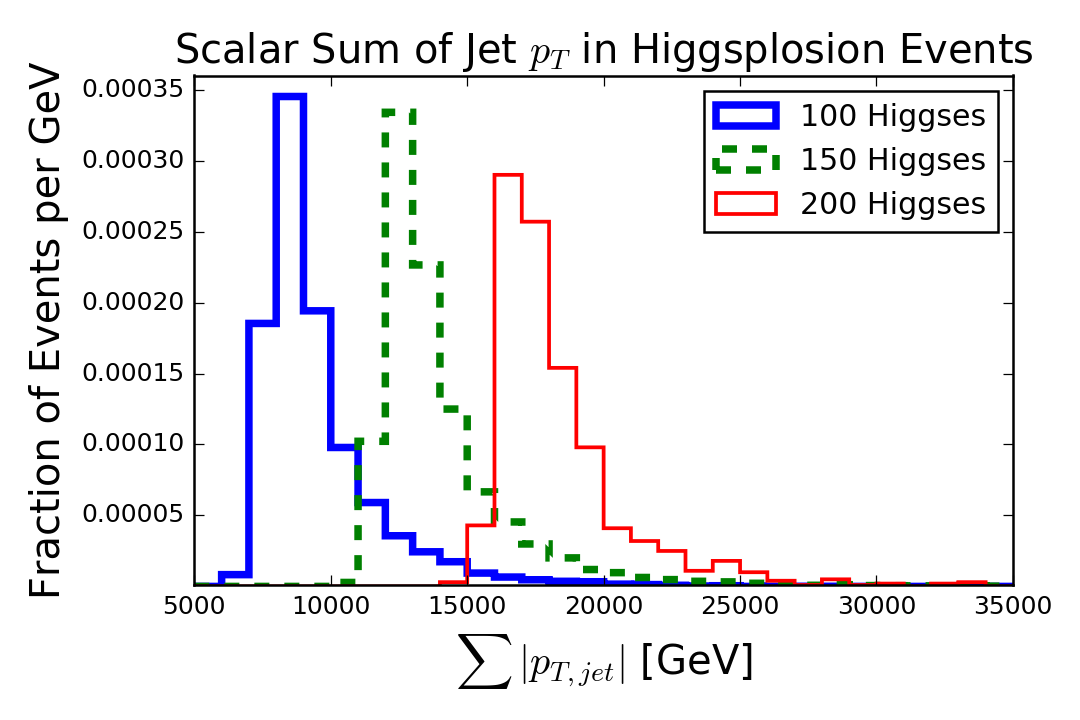}
\caption{Distribution of the scalar sum of $|p_T|$ for
all $p_T > 50$ GeV anti-$k_T$ jets
with $\Delta R < 0.4$ for $100$, $150$, and $200$
Higgs events in the final state.}
\label{fig:jet_pt}
\end{center}
\end{figure}

A more useful variable for performing the Higgs multiplicity
measurement is provided by the scalar sum of jet $p_T$,
the distributions for which are shown in Figure~\ref{fig:jet_pt}
The mean for the distribution of this variable is around $90$ GeV times the Higgs multiplicity,
while the standard deviation is between $2300 - 2700$ GeV.  Thus we can ``translate'' a measurement
of this quantity by dividing by $90$ GeV; the $1\sigma$ error for a single event is then $\lesssim 30$ events.
So a $10\%$ measurement (an error on the multiplicity of $10-20$) is possible with ten or fewer events
(treating the distributions as Gaussian, which is an approximation, but our conclusions should not be
changed by a more detailed treatment).
However reducing the error bar to the point where we are, e.g., able to distinguish a sample of
$150$ Higgs bosons from a sample of $151$ Higgs bosons at $1 \sigma$
will take $\approx 700 - 900$ events; more significant separations require more events in a straightforward
way.  It will likewise take thousands of events to determine the cross sections for various
Higgs multiplicities when events of various Higgs multiplicities are produced (as we would expect).

While Higgsplosion cross sections may be at the picobarn level or higher~\cite{Khoze:2017tjt},
making this number of events trivial to obtain, we may still want to improve
the measurement given the uncertainty on the cross sections.
For example, assuming that the LHC has not currently seen any Higgsplosion events
puts a $95\%$ confidence level limit of $\lesssim 200$ expected Higgsplosion events
in the lifetime of the $14$ LHC (assuming the LHC will collect $\approx 60$ times
the current integrated luminosity and ignoring the distinction between $13$ and $14$
TeV).  We remind the reader that Higgsplosion events are not anticipated at the $14$
TeV LHC~\cite{Khoze:2017tjt}.

\begin{figure}
    \begin{center}
        \includegraphics[width = \textwidth]{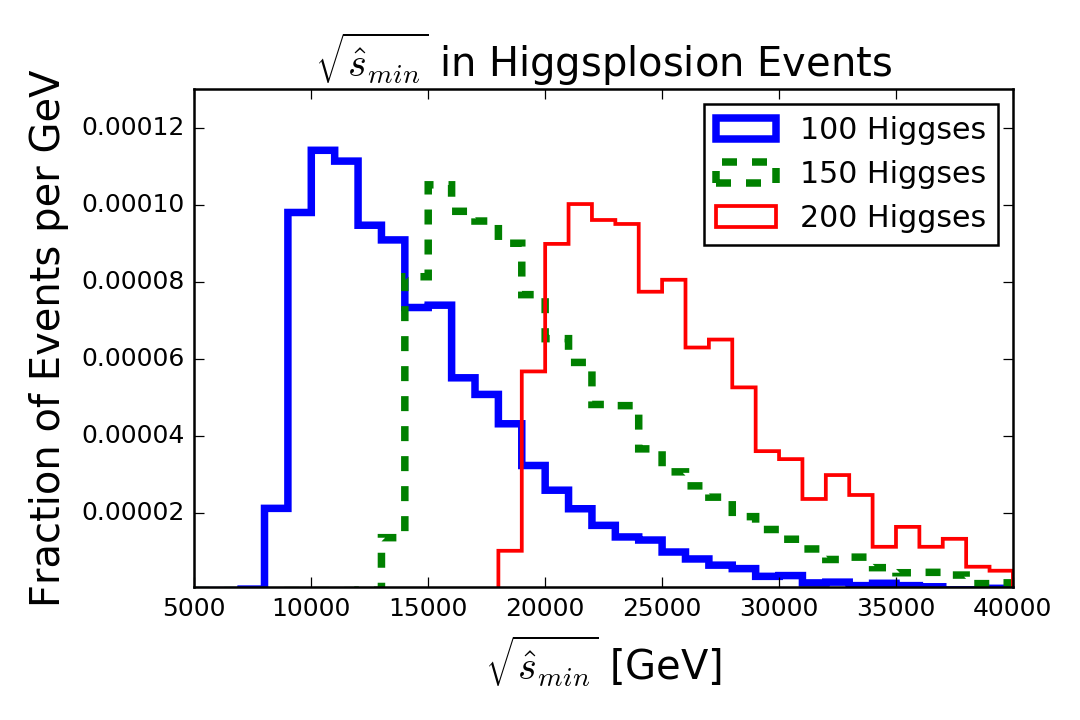}
        \caption{Distribution of $\sqrt{\hat{s}_{min}}$
                 for $100$, $150$, and $200$
                 Higgs events in the final state.}
\label{fig:smin}
    \end{center}
\end{figure}

As a first attempt to see whether more sophisticated kinematic variables
can improve on the scalar sum of jet $p_T$, we consider the $\sqrt{\hat{s}_{min}}$
variable introduced in Ref.~\cite{Konar:2008ei}, in which one finds the minimum
value of $\hat{s}$ consistent with the event if the missing energy in the event
is due to two invisible particles with a set mass.  (We set this mass to be zero in the analyses here.)
In general, this is an appealing kinematic variable, as it does not require
reconstruction of the event topology.
However, as one can see from Figure~\ref{fig:smin}, the distribution is rather broad with a long tail toward high values.
It therefore will be a less effective variable than the scalar sum of jet $p_T$ for this particular application.

\section{Conclusions}

Higgsplosion represents an interesting possibility.  Continuing theoretical research
into whether it is a valid prediction of the standard model is clearly called for.
We have found that for determining the Higgs multiplicity, the best variable,
of those we consider, is the scalar sum of jet transverse momentum.
The distributions are sufficiently narrow that a $10\%$ measurement of the
Higgs multiplicity takes only a few events, but determining the exact
multiplicity of a sample of $n$ Higgs events will take at least $\sim 1000$ events.
Similar large number of events will be required to measure $\sigma(n)$
in the case where a number of different, consecutive, values of $n$ contribute.

These analyses are therefore acceptable in the limit where picobarn cross sections for
Higgsplosion events are observed.  However, given the theoretical uncertainties, it
would be useful to be able to perform better measurements with fewer events.
Multivariate analyses may be useful toward this end, though, unfortunately
many of the variables considered here are correlated.  Substructure
variables may also be useful in improving the Higgs multiplicity measurement, as
a given jet generally includes decay products from many different Higgs bosons.

\section{Acknowledgments}
This work was supported in part by the US Department of Energy,
Office of High Energy Physics, Grant DE-SC0010504.

\section*{References}

\bibliographystyle{elsarticle-num}
\bibliography{letter}

\begin{thebibliography}{10}
\expandafter\ifx\csname url\endcsname\relax
  \def\url#1{\texttt{#1}}\fi
\expandafter\ifx\csname urlprefix\endcsname\relax\def\urlprefix{URL }\fi
\expandafter\ifx\csname href\endcsname\relax
  \def\href#1#2{#2} \def\path#1{#1}\fi

\bibitem{Cornwall:1990hh}
J.~M. Cornwall, {On the High-energy Behavior of Weakly Coupled Gauge Theories},
  Phys. Lett. B243 (1990) 271--278.
\newblock \href {http://dx.doi.org/10.1016/0370-2693(90)90850-6}
  {\path{doi:10.1016/0370-2693(90)90850-6}}.

\bibitem{Goldberg:1990qk}
H.~Goldberg, {Breakdown of perturbation theory at tree level in theories with
  scalars}, Phys. Lett. B246 (1990) 445--450.
\newblock \href {http://dx.doi.org/10.1016/0370-2693(90)90628-J}
  {\path{doi:10.1016/0370-2693(90)90628-J}}.

\bibitem{Voloshin:1992mz}
M.~B. Voloshin, {Multiparticle amplitudes at zero energy and momentum in scalar
  theory}, Nucl. Phys. B383 (1992) 233--248.
\newblock \href {http://dx.doi.org/10.1016/0550-3213(92)90678-5}
  {\path{doi:10.1016/0550-3213(92)90678-5}}.

\bibitem{Brown:1992ay}
L.~S. Brown, {Summing tree graphs at threshold}, Phys. Rev. D46 (1992)
  R4125--R4127.
\newblock \href {http://arxiv.org/abs/hep-ph/9209203}
  {\path{arXiv:hep-ph/9209203}}, \href
  {http://dx.doi.org/10.1103/PhysRevD.46.R4125}
  {\path{doi:10.1103/PhysRevD.46.R4125}}.

\bibitem{Voloshin:1992rr}
M.~B. Voloshin, {Estimate of the onset of nonperturbative particle production
  at high-energy in a scalar theory}, Phys. Lett. B293 (1992) 389--394.
\newblock \href {http://dx.doi.org/10.1016/0370-2693(92)90901-F}
  {\path{doi:10.1016/0370-2693(92)90901-F}}.

\bibitem{Voloshin:1992nu}
M.~B. Voloshin, {Summing one loop graphs at multiparticle threshold}, Phys.
  Rev. D47 (1993) R357--R361.
\newblock \href {http://arxiv.org/abs/hep-ph/9209240}
  {\path{arXiv:hep-ph/9209240}}, \href
  {http://dx.doi.org/10.1103/PhysRevD.47.R357}
  {\path{doi:10.1103/PhysRevD.47.R357}}.

\bibitem{Smith:1992rq}
B.~H. Smith, {Summing one loop graphs in a theory with broken symmetry}, Phys.
  Rev. D47 (1993) 3518--3520.
\newblock \href {http://arxiv.org/abs/hep-ph/9209287}
  {\path{arXiv:hep-ph/9209287}}, \href
  {http://dx.doi.org/10.1103/PhysRevD.47.3518}
  {\path{doi:10.1103/PhysRevD.47.3518}}.

\bibitem{Voloshin:1992xb}
M.~B. Voloshin, {Zeros of tree level amplitudes at multiboson thresholds},
  Phys. Rev. D47 (1993) 2573--2577.
\newblock \href {http://arxiv.org/abs/hep-ph/9210244}
  {\path{arXiv:hep-ph/9210244}}, \href
  {http://dx.doi.org/10.1103/PhysRevD.47.2573}
  {\path{doi:10.1103/PhysRevD.47.2573}}.

\bibitem{Argyres:1992np}
E.~N. Argyres, R.~H.~P. Kleiss, C.~G. Papadopoulos, {Amplitude estimates for
  multi - Higgs production at high-energies}, Nucl. Phys. B391 (1993) 42--56.
\newblock \href {http://dx.doi.org/10.1016/0550-3213(93)90140-K}
  {\path{doi:10.1016/0550-3213(93)90140-K}}.

\bibitem{Gorsky:1993ix}
A.~S. Gorsky, M.~B. Voloshin, {Nonperturbative production of multiboson states
  and quantum bubbles}, Phys. Rev. D48 (1993) 3843--3851.
\newblock \href {http://arxiv.org/abs/hep-ph/9305219}
  {\path{arXiv:hep-ph/9305219}}, \href
  {http://dx.doi.org/10.1103/PhysRevD.48.3843}
  {\path{doi:10.1103/PhysRevD.48.3843}}.

\bibitem{Libanov:1994ug}
M.~V. Libanov, V.~A. Rubakov, D.~T. Son, S.~V. Troitsky, {Exponentiation of
  multiparticle amplitudes in scalar theories}, Phys. Rev. D50 (1994)
  7553--7569.
\newblock \href {http://arxiv.org/abs/hep-ph/9407381}
  {\path{arXiv:hep-ph/9407381}}, \href
  {http://dx.doi.org/10.1103/PhysRevD.50.7553}
  {\path{doi:10.1103/PhysRevD.50.7553}}.

\bibitem{Voloshin:1994yp}
M.~B. Voloshin,
  \href{http://alice.cern.ch/format/showfull?sysnb=0187716}{{Nonperturbative
  methods}}, in: {27th International Conference on High-energy Physics (ICHEP
  94) Glasgow, Scotland, July 20-27, 1994}, 1994, pp. 0121--134.
\newblock \href {http://arxiv.org/abs/hep-ph/9409344}
  {\path{arXiv:hep-ph/9409344}}.
\newline\urlprefix\url{http://alice.cern.ch/format/showfull?sysnb=0187716}

\bibitem{Libanov:1995gh}
M.~V. Libanov, D.~T. Son, S.~V. Troitsky, {Exponentiation of multiparticle
  amplitudes in scalar theories. 2. Universality of the exponent}, Phys. Rev.
  D52 (1995) 3679--3687.
\newblock \href {http://arxiv.org/abs/hep-ph/9503412}
  {\path{arXiv:hep-ph/9503412}}, \href
  {http://dx.doi.org/10.1103/PhysRevD.52.3679}
  {\path{doi:10.1103/PhysRevD.52.3679}}.

\bibitem{Son:1995wz}
D.~T. Son, {Semiclassical approach for multiparticle production in scalar
  theories}, Nucl. Phys. B477 (1996) 378--406.
\newblock \href {http://arxiv.org/abs/hep-ph/9505338}
  {\path{arXiv:hep-ph/9505338}}, \href
  {http://dx.doi.org/10.1016/0550-3213(96)00386-0}
  {\path{doi:10.1016/0550-3213(96)00386-0}}.

\bibitem{Libanov:1997nt}
M.~V. Libanov, V.~A. Rubakov, S.~V. Troitsky, {Multiparticle processes and
  semiclassical analysis in bosonic field theories}, Phys. Part. Nucl. 28
  (1997) 217--240.
\newblock \href {http://dx.doi.org/10.1134/1.953038}
  {\path{doi:10.1134/1.953038}}.

\bibitem{Khoze:2014kka}
V.~V. Khoze, {Perturbative growth of high-multiplicity W, Z and Higgs
  production processes at high energies}, JHEP 03 (2015) 038.
\newblock \href {http://arxiv.org/abs/1411.2925} {\path{arXiv:1411.2925}},
  \href {http://dx.doi.org/10.1007/JHEP03(2015)038}
  {\path{doi:10.1007/JHEP03(2015)038}}.

\bibitem{Jaeckel:2014lya}
J.~Jaeckel, V.~V. Khoze, {Upper limit on the scale of new physics phenomena
  from rising cross sections in high multiplicity Higgs and vector boson
  events}, Phys. Rev. D91~(9) (2015) 093007.
\newblock \href {http://arxiv.org/abs/1411.5633} {\path{arXiv:1411.5633}},
  \href {http://dx.doi.org/10.1103/PhysRevD.91.093007}
  {\path{doi:10.1103/PhysRevD.91.093007}}.

\bibitem{Khoze:2015yba}
V.~V. Khoze, {Diagrammatic computation of multi-Higgs processes at very high
  energies: Scaling log $σ_n$ with MadGraph}, Phys. Rev. D92~(1) (2015)
  014021.
\newblock \href {http://arxiv.org/abs/1504.05023} {\path{arXiv:1504.05023}},
  \href {http://dx.doi.org/10.1103/PhysRevD.92.014021}
  {\path{doi:10.1103/PhysRevD.92.014021}}.

\bibitem{Degrande:2016oan}
C.~Degrande, V.~V. Khoze, O.~Mattelaer, {Multi-Higgs production in gluon fusion
  at 100 TeV}, Phys. Rev. D94 (2016) 085031.
\newblock \href {http://arxiv.org/abs/1605.06372} {\path{arXiv:1605.06372}},
  \href {http://dx.doi.org/10.1103/PhysRevD.94.085031}
  {\path{doi:10.1103/PhysRevD.94.085031}}.

\bibitem{Khoze:2017tjt}
V.~V. Khoze, M.~Spannowsky, {Higgsplosion: Solving the Hierarchy Problem via
  rapid decays of heavy states into multiple Higgs bosons}\href
  {http://arxiv.org/abs/1704.03447} {\path{arXiv:1704.03447}}.

\bibitem{Voloshin:2017flq}
M.~B. Voloshin, {Loops with heavy particles in multi Higgs production
  amplitudes}\href {http://arxiv.org/abs/1704.07320} {\path{arXiv:1704.07320}}.

\bibitem{Dvali:2010jz}
G.~Dvali, G.~F. Giudice, C.~Gomez, A.~Kehagias, {UV-Completion by
  Classicalization}, JHEP 08 (2011) 108.
\newblock \href {http://arxiv.org/abs/1010.1415} {\path{arXiv:1010.1415}},
  \href {http://dx.doi.org/10.1007/JHEP08(2011)108}
  {\path{doi:10.1007/JHEP08(2011)108}}.

\bibitem{Dvali:2010ns}
G.~Dvali, D.~Pirtskhalava, {Dynamics of Unitarization by Classicalization},
  Phys. Lett. B699 (2011) 78--86.
\newblock \href {http://arxiv.org/abs/1011.0114} {\path{arXiv:1011.0114}},
  \href {http://dx.doi.org/10.1016/j.physletb.2011.03.054}
  {\path{doi:10.1016/j.physletb.2011.03.054}}.

\bibitem{Dvali:2016ovn}
G.~Dvali,
  \href{http://inspirehep.net/record/1477803/files/arXiv:1607.07422.pdf}{{Strong
  Coupling and Classicalization}}, in: {Proceedings, LHCSki 2016 - A First
  Discussion of 13 TeV Results: Obergurgl, Austria, April 10-15, 2016}, 2016.
\newblock \href {http://arxiv.org/abs/1607.07422} {\path{arXiv:1607.07422}}.
\newline\urlprefix\url{http://inspirehep.net/record/1477803/files/arXiv:1607.07422.pdf}

\bibitem{Alwall:2006yp}
J.~Alwall, et~al., {A Standard format for Les Houches event files}, Comput.
  Phys. Commun. 176 (2007) 300--304.
\newblock \href {http://arxiv.org/abs/hep-ph/0609017}
  {\path{arXiv:hep-ph/0609017}}, \href
  {http://dx.doi.org/10.1016/j.cpc.2006.11.010}
  {\path{doi:10.1016/j.cpc.2006.11.010}}.

\bibitem{Ball:2014uwa}
R.~D. Ball, et~al., {Parton distributions for the LHC Run II}, JHEP 04 (2015)
  040.
\newblock \href {http://arxiv.org/abs/1410.8849} {\path{arXiv:1410.8849}},
  \href {http://dx.doi.org/10.1007/JHEP04(2015)040}
  {\path{doi:10.1007/JHEP04(2015)040}}.

\bibitem{Buckley:2014ana}
A.~Buckley, J.~Ferrando, S.~Lloyd, K.~Nordström, B.~Page, M.~Rüfenacht,
  M.~Schönherr, G.~Watt, {LHAPDF6: parton density access in the LHC precision
  era}, Eur. Phys. J. C75 (2015) 132.
\newblock \href {http://arxiv.org/abs/1412.7420} {\path{arXiv:1412.7420}},
  \href {http://dx.doi.org/10.1140/epjc/s10052-015-3318-8}
  {\path{doi:10.1140/epjc/s10052-015-3318-8}}.

\bibitem{Sjostrand:2014zea}
T.~Sjöstrand, S.~Ask, J.~R. Christiansen, R.~Corke, N.~Desai, P.~Ilten,
  S.~Mrenna, S.~Prestel, C.~O. Rasmussen, P.~Z. Skands, {An Introduction to
  PYTHIA 8.2}, Comput. Phys. Commun. 191 (2015) 159--177.
\newblock \href {http://arxiv.org/abs/1410.3012} {\path{arXiv:1410.3012}},
  \href {http://dx.doi.org/10.1016/j.cpc.2015.01.024}
  {\path{doi:10.1016/j.cpc.2015.01.024}}.

\bibitem{Dobbs:2001ck}
M.~Dobbs, J.~B. Hansen, {The HepMC C++ Monte Carlo event record for High Energy
  Physics}, Comput. Phys. Commun. 134 (2001) 41--46.
\newblock \href {http://dx.doi.org/10.1016/S0010-4655(00)00189-2}
  {\path{doi:10.1016/S0010-4655(00)00189-2}}.

\bibitem{Cacciari:2011ma}
M.~Cacciari, G.~P. Salam, G.~Soyez, {FastJet User Manual}, Eur. Phys. J. C72
  (2012) 1896.
\newblock \href {http://arxiv.org/abs/1111.6097} {\path{arXiv:1111.6097}},
  \href {http://dx.doi.org/10.1140/epjc/s10052-012-1896-2}
  {\path{doi:10.1140/epjc/s10052-012-1896-2}}.

\bibitem{Baer:2016wkz}
H.~Baer, V.~Barger, J.~S. Gainer, P.~Huang, M.~Savoy, D.~Sengupta, X.~Tata,
  {Gluino reach and mass extraction at the LHC in radiatively-driven natural
  SUSY}\href {http://arxiv.org/abs/1612.00795} {\path{arXiv:1612.00795}}.

\bibitem{Konar:2008ei}
P.~Konar, K.~Kong, K.~T. Matchev, {$\sqrt{\hat{s}}_{min}$ : A Global inclusive
  variable for determining the mass scale of new physics in events with missing
  energy at hadron colliders}, JHEP 03 (2009) 085.
\newblock \href {http://arxiv.org/abs/0812.1042} {\path{arXiv:0812.1042}},
  \href {http://dx.doi.org/10.1088/1126-6708/2009/03/085}
  {\path{doi:10.1088/1126-6708/2009/03/085}}.

\end{thebibliography}

\end{document}